\documentclass[aps,prb,reprint,twocolumn,superscriptaddress,floatfix,10pt,letterpaper,showpacs]{revtex4-1}
\usepackage{amsmath,amssymb,amsfonts,stmaryrd,wasysym,graphicx,multirow,color,textcomp}
\usepackage{url}
\usepackage[colorlinks=true,citecolor=blue,urlcolor=blue]{hyperref}

\newcommand{\TI}{\hyperlink{TI}{TI} }
\newcommand{\FTI}{\hyperlink{FTI}{FTI} }
\newcommand{\TR}{\hyperlink{TR}{TR} }
\newcommand{\QP}{\hyperlink{QP}{QP} }
\newcommand{\TO}{\hyperlink{TO}{TO} }
\newcommand{\FS}{\hyperlink{FS}{FS} }
\newcommand{\SCS}{\hyperlink{SCS}{SCS} }
\newcommand{\TPf}{\hyperlink{TPf}{$\mathcal{T}$-$\mathrm{Pf}^\ast$} }
\newcommand{\CFT}{\hyperlink{CFT}{CFT} }
\newcommand{\FQH}{\hyperlink{FQH}{FQH} }
\newcommand{\MBS}{\hyperlink{MBS}{MBS} }
\newcommand{\PH}{\hyperlink{PH}{PH} }
\newcommand{\LLL}{\hyperlink{LLL}{LLL} }

\begin{document}

\title{Surfaces and slabs of fractional topological insulator heterostructures}
\author{Sharmistha Sahoo}\affiliation{Department of Physics, University of Virginia, Charlottesville, VA22904, USA}
\author{Alexander Sirota}\affiliation{Department of Physics, University of Virginia, Charlottesville, VA22904, USA}
\author{Gil Young Cho}\affiliation{School of Physics, Korea Institute for Advanced Study, Seoul 02455, Korea}
\author{Jeffrey C. Y. Teo}\email{jteo@virginia.edu}\affiliation{Department of Physics, University of Virginia, Charlottesville, VA22904, USA}

\begin{abstract}
Fractional topological insulators (FTI) are electronic topological phases in $(3+1)$ dimensions enriched by time reversal (TR) and charge $U(1)$ conservation symmetries. We focus on the simplest series of fermionic FTI, whose bulk quasiparticles consist of deconfined partons that carry fractional electric charges in integral units of $e^\ast=e/(2n+1)$ and couple to a discrete $\mathbb{Z}_{2n+1}$ gauge theory. We propose massive symmetry preserving or breaking FTI surface states. Combining the long-ranged entangled bulk with these topological surface states, we deduce the novel topological order of quasi-$(2+1)$ dimensional FTI slabs as well as their corresponding edge conformal field theories.
\end{abstract}

\maketitle

\section{\label{sec:Intro}Introduction}

Conventional topological insulators (\hypertarget{TI}{TI})~\cite{FuKaneMele3D,Roy07,MooreBalents07,QiHughesZhang08} are time reversal (\hypertarget{TR}{TR}) and charge $U(1)$ symmetric electronic band insulators in three dimensions that host massless surface Dirac fermions. The topologically protected surface Dirac fermion can acquire a single-body ferromagnetic or superconducting mass by breaking \TR or charge $U(1)$ symmetry respectively. Alternatively it can acquire a many-body interacting mass while preserving both symmetries, and exhibit long-ranged entangled surface topological order~\cite{WangPotterSenthilgapTI13,MetlitskiKaneFisher13b,ChenFidkowskiVishwanath14,BondersonNayakQi13}. 
On the other hand, fractional topological insulators (\hypertarget{FTI}{FTI})~\cite{MaciejkoQiKarchZhang10,SwingleBarkeshliMcGreevySenthil11,maciejko2012models,ye2016composite,maciejko2015fractionalized,stern2016fractional,YeChengFradkin17} are long-range entangled topologically ordered electronic phases in $(3+1)$ dimensions outside of the single-body mean-field band theory description. 
They carry \TR and charge $U(1)$ symmetries, which enrich its topological order (\hypertarget{TO}{TO}) in the sense that a symmetric surface must be anomalous and cannot be realized non-holographically by a true $(2+1)$-D system. In this Rapid Communication, we describe the topological properties of various massive surface states and quasi-$(2+1)$-D slabs of a series of \FTI. In particular, we focus on the quasi-particle (\hypertarget{QP}{QP}) structure.


\begin{figure}[htbp]
\centering\includegraphics[width=0.5\textwidth]{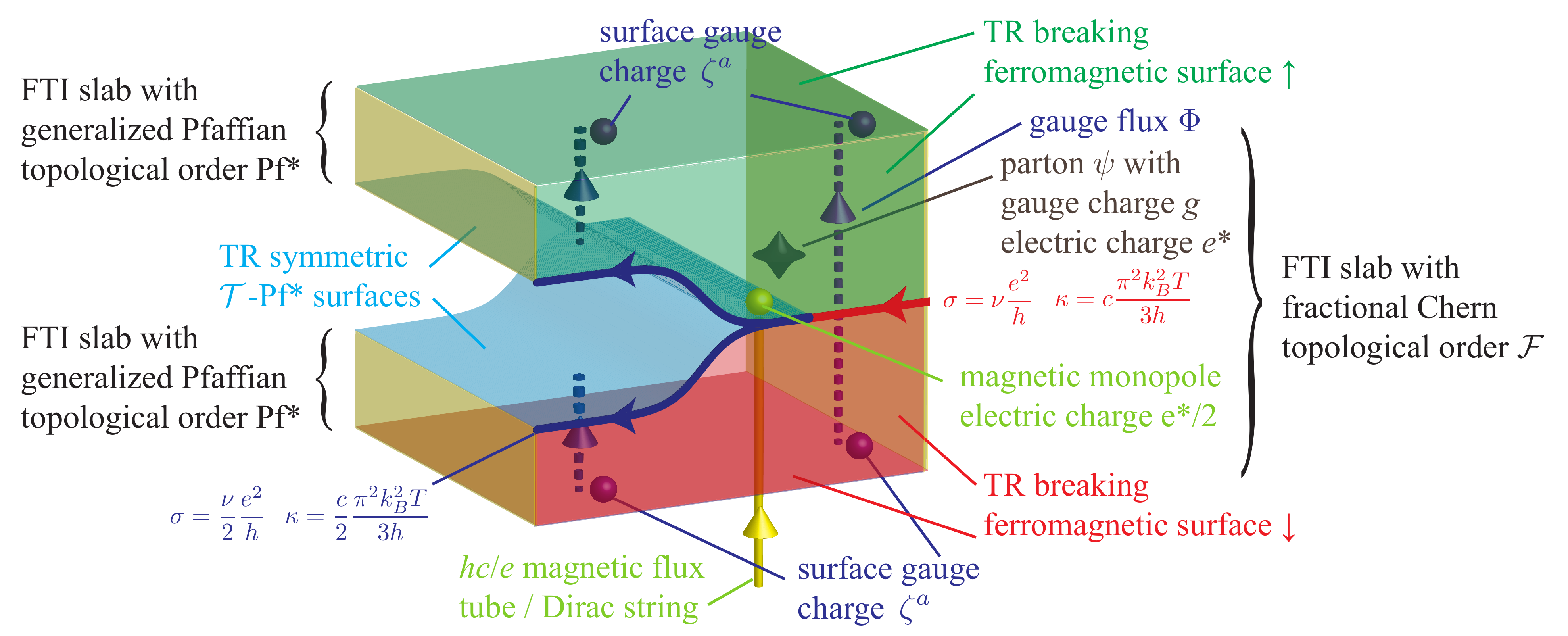}
\caption{Summary of the QP and gauge flux content in FTI slabs. A pair of $\mathrm{Pf}^\ast$ FTI slabs are merged into a fractional Chern FTI slab $\mathcal{F}$ by gluing the two TR symmetric $\mathcal{T}$-$\mathrm{Pf}^\ast$ surfaces. Directed bold lines on the front surface are chiral edge modes of the $\mathrm{Pf}^\ast$ and $\mathcal{F}$ FTI slabs.}\label{fig1}
\end{figure}

We focus on a series of fermionic \FTI, labeled by integers $n$, whose magneto-electric response is characterized by the $\theta$-angle $\theta=\pi/(2n+1)$ (modulo $2\pi/(2n+1)$) that associates an electric charge of $e^\ast/2=e/2(2n+1)$ to each magnetic monopole~\cite{Witten79}, for $e$ the electric charge of the electron. In particular, we consider \FTI that support deconfined fermionic parton excitations $\psi$ in the bulk, each carrying a fractional electric charge of $e^\ast=e/(2n+1)$. The electronic \QP decomposes as $\psi_{\mathrm{el}}\sim\psi_1\ldots\psi_{2n+1}$. The $(3+1)$-D \TO is based on a discrete $\mathbb{Z}_{2n+1}$ gauge theory~\cite{maciejko2012models}. The theory supports electrically neutral string-like gauge flux $\Phi$, so that a monodromy quantum phase of $e^{2\pi ig/(2n+1)}$ is obtained each time $\psi$ orbits around it. In other words, $\psi$ carries the gauge charge $g$. The integer $g$ and $2n+1$ are relatively prime so that all local \QP must be combinations of the electronic \QP $\psi_{\mathrm{el}}$ and must carry integral electric charges and trivial gauge charges.

Generalizing the surface state of a conventional \TI, the surface of a \FTI hosts massless Dirac partons coupling with a $\mathbb{Z}_{2n+1}$ gauge theory. Unlike its non-interacting counterpart whose gapless Dirac surface state is symmetry protected in the single-body picture, a \FTI is strongly interacting to begin with and there is no topological reason for its surface state to remain gapless. In this Rapid Communication we focus on three types of gapped surface states -- ferromagnetic surfaces (\hypertarget{FS}{FS}) that break \TR, superconducting surfaces (\hypertarget{SCS}{SCS}) that break charge $U(1)$, and symmetric surfaces which generalize the $\mathcal{T}$-Pfaffian surface state of a conventional \TI and is denoted by \hypertarget{TPf}{$\mathcal{T}$-$\mathrm{Pf}^\ast$}. The topological order for \FTI slab with these surfaces are discussed in Sec.~\ref{FS}, \ref{SCS} and \ref{TPF} respectively. In Sec.~\ref{Gluing}, we discuss, using an anyon condensation picture, the gluing of a pair of $\mathcal{T}$-Pfaffian surfaces. We conclude in Sec.~\ref{Conclusion} with remarks on a complementary way to understand these topological order \cite{ChoTeoFradkin17}.

\section{\label{FS}Ferromagetic Heterostructure}
We begin with a slab that has opposite \TR breaking \FS. In the \FS, in addition to the single-body Dirac mass $m$ for the surface parton, the $\mathbb{Z}_{2n+1}$ gauge sector also shows \TR breaking signature. The $\mathbb{Z}_{2n+1}$ gauge theory is only present inside the \FTI, and when a flux line $\Phi$ terminates at the surface, the \TR breaking boundary condition confines an electrically neutral surface gauge \QP, denoted by $\zeta^a$, with gauge charge $a$ at the flux-surface junction (see Fig.~\ref{fig1}). This gauge flux-charge composite, referred to as a dyon $\delta=\Phi\times\zeta^a$, carries fractional spin $h_\delta=a/(2n+1)$ because a $2\pi$-rotation about the normal axis braids $a$ gauge charges around $\Phi$ and results in the monodromy quantum phase of $e^{2\pi ia/(2n+1)}$. \TR conjugates all quantum phases so, $a\not\equiv0$ modulo $2n+1$ breaks \TR.

The one-dimensional interface between two \TR conjugate \FS domains hosts a fractional chiral channel. For example, the interface between two \FS domains with opposite ferromagnetic orientations on the surface of a conventional \TI bounds a chiral Dirac channel~\cite{TeoKane,FuKanechargetransport09,QiWittenZhang13}, where electrons propagate only in the forward direction. Alternatively, a \TI slab with opposite \TR breaking \FS is topologically identical to a quasi-$(2+1)$-D Chern insulator~\cite{Haldane1988,liu2016quantum} and supports a chiral Dirac edge mode. Similarly, in the \FTI case, the low-energy content of the fractional chiral channel between a pair of \TR conjugate \FS domains can be inferred by the edge mode of a \FTI slab with \TR breaking \FS that is topologically identical to a quasi-$(2+1)$-D fractional Chern insulator~\cite{RegnaultBernevigfractionchern,NeupertSantosChamonMudry11,TangMeiWen11,ShengGuSunSheng11} or fractional quantum Hall (\hypertarget{FQH}{FQH}) state~\cite{FQHE_Review}. The chiral $(1+1)$-D channel is characterized by two response quantities~\cite{Laughlin_IQHE, Halperin82, Hatsugai93, Schulz00, Volovik92, KaneFisher97, Cappelli01, Kitaev06, Luttinger64} -- the differential electric conductance $\sigma=dI/dV=\nu e^2/h$ that relates the changes of electric current and potential, and the differential thermal conductance $\kappa=dI_T/dT=c(\pi^2k_B^2/3h)T$ that relates the changes of energy current and temperature. In the slab geometry, they are equivalent to the Hall conductance $\sigma=\sigma_{xy}$, $\kappa=\kappa_{xy}$. $\nu=N_e/N_\phi$ is also referred to as the filling fraction of the \FTI slab and associates the gain of electric charge (in units of $e$) to the addition of a magnetic flux quantum $hc/e$. $c=c_R-c_L$ is the chiral central charge of the conformal field theory (\hypertarget{CFT}{CFT})~\cite{bigyellowbook} that effectively describes the low-energy degrees of freedom of the fractional chiral channel. 

Since the top and bottom surfaces of the \FTI slab are \TR conjugate, their parton Dirac masses $m$ and gauge flux-charge ratio $a$ have opposite signs. The anyon content is generated by the partons and gauge dyons. When a gauge flux passes through the entire slab geometry from the bottom to the top surface, it associates with total $2a$ gauge charges at the two surface junctions. We denote this dyon by $\gamma=\Phi\times\zeta^{2a}$, which corresponds to an electrically neutral anyon in the slab with spin $h_\gamma=2a/(2n+1)$. If $a$ is relatively prime with $2n+1$, the primitive dyon generates the chiral Abelian topological field theory $\mathbb{Z}_{2n+1}^{(2a)}$~\cite{MooreSeiberg89,Bondersonthesis}, which consists of the dyons $\gamma^m$, for $m=0,\ldots,2n$, with spins $h_{\gamma^m}=2am^2/(2n+1)$ modulo 1 and fusion rules $\gamma^m\times\gamma^{m'}=\gamma^{m+m'}$, $\gamma^{2n+1}=\gamma^0=1$. In particular, when $a=-1$, $\gamma^n$ now has spin $-2n^2/(2n+1)\equiv n/(2n+1)$ modulo 1, which is identical to that of the fundamental \QP of the $SU(2n+1)$ Chern-Simons theory at level 1~\cite{MooreSeiberg89,Bondersonthesis}. This identifies the Abelian theories $\mathbb{Z}_{2n+1}^{(-2)}\cong\mathbb{Z}_{2n+1}^{(n)}=SU(2n+1)_1$, which has chiral central charge $c_{\mathrm{neutral}}=2n$.

The \FTI slab also supports fractionally charged partons $\psi$, each carrying a gauge charge $g$. The electrically charged sector can be decoupled from the neutral $\mathbb{Z}_{2n+1}^{(2a)}$ sector by combining each parton with a specific number of dyons $\lambda=\psi\times\gamma^{-n^2ug}$, where $ua+v(2n+1)=1$ for some integer $u$, $v$, so that the combination is local (i.e.~braids trivially) with any dyons $\gamma^m$. $\lambda$ has fractional electric charge $q_\lambda=e^\ast$ and spin $h_\lambda=1/2+n^3ug^2/(2n+1)$ modulo 1. The $\langle\mathrm{charge}\rangle$ sector consists of the fractional Abelian \QP products $\lambda^m$, where $\lambda^{2n+1}\sim\psi^{2n+1}\sim\psi_{\mathrm{el}}$ corresponds to the local electronic \QP. In particular, when $a=-1$ and $g=-2$, $h_\lambda=1/2(2n+1)$ and therefore $\lambda$ behaves exactly like the Laughlin \QP of the \FQH state $U(1)_{(2n+1)/2}$ with filling fraction $\nu=1/(2n+1)$ and chiral central charge $c_{\mathrm{charge}}=1$. 
Combining the neutral and charge sectors, the \FTI slab with \TR breaking \FS has the decoupled tensor product \TO \begin{align}\mathcal{F}=\langle\mathrm{charge}\rangle\otimes\mathbb{Z}_{2n+1}^{(2a)},\label{FTIFSFS}\end{align} and in the special case when $a=-1$ and $g=-2$, it is identical to the Abelian state $U(1)_{(2n+1)/2}\otimes SU(2n+1)_1$, which has a total central charge $c=2n+1$. In general, the filling fraction and chiral central charge are not definite and are subject to surface reconstruction. For instance, the addition of $2N$ electronic Dirac fermions per surface modifies the two response quantities by an equal amount $\nu\to\nu+2N$, $c\to c+2N$.
\section{\label{SCS}Superconducting Heterostructure}
Next we move on to superconducting heterostructures. We begin with the fractional Chern \FTI slab $\mathcal{F}$ in \eqref{FTIFSFS} and introduce weak superconducting pairing, perhaps induced by proximity with a bulk superconductor, without closing the bulk energy gap. In the simplest scenario, this condenses all parton pairs $\psi^{2m}$, which form a {\em Lagrangian subgroup}~\cite{Levin13} -- a maximal set of mutually local bosons -- containing the Cooper pair $\psi_{\mathrm{el}}^2=\psi^{2(2n+1)}$. Since the parton pair $\psi^2$ carries gauge charge $2g$, which is relatively prime with $2n+1$, the condensate confines all non-trivial dyons $\gamma^m$, which are non-local and have non-trivial monodromy with $\psi^2$. As the neutral sector $\mathbb{Z}_{2n+1}^{(2a)}$ is killed by pairing, the superconducting \FTI slab with \TR conjugate \FS has a trivial fermionic \TO. It however still carries chiral fermionic edge modes with the same chiral central charge $c_{\mathcal{F}}$. On the other hand, these fermionic channels also live along the line interface between \TR conjugate ferromagnetic domains on the surface of a weakly superconducting \FTI. When the line interface hits a \TR symmetric \SCS island (c.f.~ Fig.~\ref{fig1} by replacing the \TPf surfaces by \SCS), these chiral channels split and divide along the pair of \SCS-\FS line interfaces. Both of these channels are electrically neutral as charge $U(1)$ symmetry is broken by the superconductor, and each of them carries half of the energy current of $\mathcal{F}$ and has chiral central charge $c_{\mathcal{F}}/2$. For example, the \SCS-\FS heterostructure on a conventional \TI surface holds a chiral Majorana channel with $c=1/2$ along the line tri-junction~\cite{FuKanechargetransport09,TeoKane}. In the specific fractional case when $a=-1$ and $g=-2$, each \SCS-\FS line interfaces holds $2n+1$ chiral Majorana fermions and is described by the Wess-Zumino-Witten $SO(2n+1)_1$ \CFT with the central charge $c=(2n+1)/2$. Analogous to the conventional superconducting \TI surface~\cite{FuKane08}, the \SCS of the \FTI supports a zero energy Majorana bound state (\hypertarget{MBS}{MBS}) at a vortex core. Now that the condensate consists of parton pairs, vortices are quantized with the magnetic flux $hc/2e^\ast=(2n+1)hc/2e$. Each pair of \MBS forms a two-level system distinguished by parton fermion parity.

\section{\label{TPF}Generalized \texorpdfstring{$\mathcal{T}$}{T}-Pfaffian* surface state}
Lastly, we describe the \TPf surface state that preserves both \TR and charge $U(1)$ symmetries of the FTI. Generalizing the $\mathcal{T}$-Pfaffian symmetric gapped surface state of a conventional \TI described in Ref.\cite{ChenFidkowskiVishwanath14}, the \FTI version -- referred here as $\mathcal{T}$-Pfaffian$^\ast$ -- consists of the Abelian surface anyons $\openone_j$ and $\Psi_j$, for $j$ even, and the non-Abelian Ising-like anyons $\Sigma_j$, for $j$ odd. The index $j$ corresponds to the fractional electric charge $q_j=je/4(2n+1)$. The surface anyons satisfy the fusion rules \begin{gather}\openone_j\times\openone_{j'}=\Psi_j\times\Psi_{j'}=\openone_{j+j'},\quad\openone_j\times\Psi_{j'}=\Psi_{j+j'},\nonumber\\\Psi_j\times\Sigma_{j'}=\Sigma_{j+j'},\quad\Sigma_j\times\Sigma_{j'}=\openone_{j+j'}+\Psi_{j+j'},\label{TPffusion}\end{gather} and the spin statistics \begin{gather}h_{\openone_j}=h_{\Psi_j}-\frac{1}{2}=\frac{j^2}{16},\quad h_{\Sigma_j}=\frac{j^2-1}{16}\quad\mbox{modulo 1}\end{gather} so that $\openone_j,\Psi_j$ are bosonic, fermionic or semionic, and $\Sigma_j$ are bosonic or fermionic. The fermion $\Psi_4$ is identical to the super-selection sector of the bulk parton $\psi$, which is local with respect to all surface anyons and can escape from the surface and move into the bulk. \TR symmetry acts on the surface anyons the same way it acts on those in the $\mathcal{T}$-Pfaffian state for conventional \TI~\cite{ChenFidkowskiVishwanath14,ChoTeoFradkin17}. For example, the parton combinations $\psi^{2j+1}=\Psi_{8j+4}$ (and $\psi^{2j}=\openone_{8j}$) are Kramers doublet fermions (respectively Kramers singlet bosons), while $\Psi_{8j}$ ($\openone_{8j+4}$) are Kramers singlet fermions (respectively Kramers doublet bosons). Moreover, for identical reasons as in the conventional \TI case, the \TPf state is anomalous and can only be supported holographically on the surface of a topological bulk. For instance, the bosonic \TO of the \TPf state after gauging fermion parity would necessarily 
violate \TR symmetry. We notice in passing that there are alternative surface \TO that generalize those in Refs.\cite{WangPotterSenthilgapTI13,MetlitskiKaneFisher13b}. However we will only focus on the \TPf state in this Rapid Communication.

The \FTI slab with a \TR symmetric \TPf top surface and a \TR breaking bottom \FS carries a novel quasi-$(2+1)$-D \TO. Its topological content consists of the fractional partons coupled with the $\mathbb{Z}_{2n+1}$ gauge theory in the bulk and the \TPf surface state (see Fig.~\ref{fig1}). All surface anyons are confined to the \TR symmetric surface except the parton combinations $\psi^{2j+1}=\Psi_{8j+4}$ and $\psi^{2j}=\openone_{8j}$. Moreover, the \TR breaking boundary condition confines a gauge \QP $\zeta^a$ per gauge flux $\Phi$ ending on the \FS. On the other hand, there is no gauge charge associated with a gauge flux ending on the \TPf surface because of \TR symmetry. Thus a gauge flux passing through the entire slab corresponds to the dyon $\delta=\Phi\times\zeta^a$ with spin $h_\delta=a/(2n+1)$ modulo 1. The \TPf state couples non-trivially to the $\mathbb{Z}_{2n+1}$ gauge theory as the parton $\psi=\Psi_4$ carries a gauge charge $g$. The general surface anyons $X_j$, for $X=\openone,\Psi,\Sigma$, must carry the gauge charge $z(j)\equiv n^2gj$ modulo $2n+1$ and associate to the monodromy quantum phase $e^{2\pi iz(j)/(2n+1)}$ when orbiting around the dyon $\delta$. For instance, as $2n\equiv-1$ modulo $2n+1$, $z(4j)\equiv gj$ counts the gauge charge of the parton combination $\psi^j$.

The \TO of this \FTI slab is therefore generated by combinations of the \TPf anyons and the dyon $\delta$. We denote the composite anyon by \begin{align}\tilde{X}_{j,z}=X_j\otimes\delta^{z+n^3ugj},\label{Zfanyon}\end{align} where $X=\openone,\Psi$ for $j$ even or $\Sigma$ for $j$ odd, $z=0,\ldots,2n$ modulo $2n+1$, and $ua+v(2n+1)=1$. They satisfy the fusion rules \begin{gather}\tilde\openone_{j,z}\times\tilde\openone_{j',z'}=\tilde\Psi_{j,z}\times\tilde\Psi_{j',z'}=\tilde\openone_{j+j',z+z'},\nonumber\\\tilde\openone_{j,z}\times\tilde\Psi_{j',z'}=\tilde\Psi_{j+j',z+z'},\quad\tilde\Psi_{j,z}\times\tilde\Sigma_{j',z'}=\tilde\Sigma_{j+j',z+z'},\nonumber\\\tilde\Sigma_{j,z}\times\tilde\Sigma_{j',z'}=\tilde\openone_{j+j',z+z'}+\tilde\Psi_{j+j',z+z'}.\label{Zffusion}\end{gather} They follow the spin statistics \begin{align}h(\tilde\openone_{j,z})&=h(\tilde\Psi_{j,z})-\frac{1}{2}=h(\tilde\Sigma_{j,z})+\frac{1}{16}\nonumber\\&=\frac{j^2}{16}+\frac{az^2-n^6ug^2j^2}{2n+1}\quad\mbox{modulo 1}.\end{align} The $j,z$ indices in \eqref{Zfanyon} are defined in a way so that $\tilde{X}_{j,0}$ are local with respect to the dyons $\delta^z=\tilde\openone_{0,z}$ and decoupled from the dyon sector $\mathbb{Z}_{2n+1}^{(a)}$. The \TPf surface anyons belong to the subset $X_j=\tilde{X}_{j,-n^3ugj}$, which is a maximal sub-category that admits a \TR symmetry. The electronic \QP belongs to the super-selection sector $\psi_{\mathrm{el}}=\tilde\Psi_{4(2n+1),0}$, which is local with respect to all anyons. If one gauges fermion parity and includes anyons that associate $-1$ monodromy phase with $\psi_{\mathrm{el}}$, i.e.~if one includes $\tilde\openone_{j,z},\tilde\Psi_{j,z}$ for $j$ odd and $\tilde\Sigma_{j,z}$ for $j$ even, the $\langle\overline{\mathrm{Ising}}\rangle$ sector generated by $1=\tilde\openone_{0,0}$, $f=\tilde\Psi_{0,0}$, $\sigma=\tilde\Sigma_{0,0}$ is local with and decoupled from the $\langle\mathrm{charge}\rangle_{\mathrm{Pf}^\ast}$ sector generated by $\tilde\openone_{j,0}$. The \TO of the \FTI slab thus takes the decoupled tensor product form after gauging fermion parity \begin{align}\mathrm{Pf}^\ast=\langle\mathrm{charge}\rangle_{\mathrm{Pf}^\ast}\otimes\langle\overline{\mathrm{Ising}}\rangle\otimes\mathbb{Z}_{2n+1}^{(a)}.\label{ZfTO}\end{align} Gauging fermion parity is not the focus of this Rapid Communication. Nevertheless, we notice in passing that there are inequivalent ways of fermion parity gauging, and in order for the $\mathrm{Pf}^\ast$ theory to have the appropriate central charge, \eqref{ZfTO} needs to be modified by a neutral Abelian $SO(2n)_1$ sector~\cite{ChoTeoFradkin17}. However, the tensor product \eqref{ZfTO} is sufficient and correct to describe the fermionic \TO of the \FTI slab (with global ungauged fermion parity) by restricting to super-selection sectors $\tilde{X}_{j,z}$ that are local with respect to the electronic \QP $\psi_{\mathrm{el}}$. We refer to this fermionic \TO as a generalized Pfaffian state.

\section{\label{Gluing}Gluing T-Pfaffian* surfaces}
The chiral channel $\mathcal{F}$ in \eqref{FTIFSFS} between a pair of \TR conjugate \FS domains divides into a pair of fermionic $\mathrm{Pf}^\ast$ in \eqref{ZfTO} at a junction where the two \FS domains sandwich a \TR symmetric \TPf surface domain (see Fig.~\ref{fig1}). Conservation of charge and energy requires the filling fractions and chiral central charges to equally split, i.e.~$2\nu_{\mathrm{Pf}^\ast}=\nu_{\mathcal{F}}$ and $2c_{\mathrm{Pf}^\ast}=c_{\mathcal{F}}$. For instance, in the prototype case when $a=-1$ and $g=-2$, $\nu_{\mathrm{Pf}^\ast}=1/2(2n+1)$ and $c_{\mathrm{Pf}^\ast}=(2n+1)/2$. Similar to the aforementioned $\mathcal{F}$ case, these quantities are subjected to surface reconstruction $\nu\to\nu+N$, $c\to c+N$. 

In addition to the response quantities, the \TO of $\mathcal{F}$ for the \FTI slab with \TR conjugate \FS is related to that of the fermionic $\mathrm{Pf}^\ast$ by a {\em relative tensor product} \begin{align}\mathcal{F}=\mathrm{Pf}^\ast\boxtimes_b\mathrm{Pf}^\ast.\label{gluing}\end{align} This can be understood by juxtaposing the \TR symmetric surfaces of a pair of $\mathrm{Pf}^\ast$ \FTI slabs and condensing surface bosonic anyon pairs on the two \TPf surfaces. This anyon condensation~\cite{BaisSlingerlandCondensation,Kong14,NeupertHeKeyserlingkSierraBernevig16} procedure effectively glues the two \FTI slabs together along the \TR symmetric surfaces  (see Fig.~\ref{fig1}). The relative tensor product $\boxtimes_b$ involves first taking a decoupled tensor product $\otimes$ when the two $\mathrm{Pf}^\ast$ \FTI slabs are put side by side. 
Among the \TR symmetric surface anyons in $(\mathcal{T}$-$\mathrm{Pf}^\ast)^A\otimes(\mathcal{T}$-$\mathrm{Pf}^\ast)^B$ where $A,B$ refers to the two slabs, we condense the collection of electrically neutral bosonic pairs \begin{align}b=\left\{\begin{array}{*{20}c}\openone^A_{4j}\openone^B_{-4j},\Psi^A_{4j}\Psi^B_{-4j},\openone^A_{4j+2}\Psi^B_{-4j-2},\\\Psi^A_{4j+2}\openone^B_{-4j-2},\Sigma^A_{2j+1}\Sigma^B_{-2j-1}\end{array}\right\}.\label{bosons}\end{align} All anyons that are non-local with respect to and braid non-trivially around any of the bosons in $b$ are confined. This includes all anyon combinations $\tilde{X}^A_{j_a,z_a}\tilde{X}^B_{j_b,z_b}$ where the dyon numbers $z_a+n^3ugj_a$ and $z_b+n^3ugj_b$ disagree modulo $2n+1$. Physically, this ensures gauge fluxes must continue through both $A$ and $B$ slabs, or equivalently all gauge monopoles at the interface are confined as they signify imbalances of gauge fluxes through the two slabs. The new deconfined dyon $\gamma=\tilde\openone^A_{0,1}\tilde\openone^B_{0,1}$ consists of a gauge flux that passes continuously across both slabs with gauge \QP $\zeta^a$ on each of the remaining top and bottom \TR breaking surfaces. A deconfined anyon thus splits into a dyon component $\gamma^z$ and a surface component in $(\mathcal{T}$-$\mathrm{Pf}^\ast)^A\otimes(\mathcal{T}$-$\mathrm{Pf}^\ast)^B$. Within the surface part, all combinations that involve only $\Sigma^A$ or only $\Sigma^B$ are confined by the $\Psi^A_0\Psi^B_0$ condensate. Other confined anyons include $\openone^A_{j_a}\openone^B_{j_b}$, $\Psi^A_{j_a}\Psi^B_{j_b}$, $\openone^A_{j_a+2}\Psi^B_{j_b-2}$, $\Psi^A_{j_a+2}\openone^B_{j_b-2}$ and $\Sigma^A_{j_a\pm1}\Sigma^B_{j_b\mp1}$ for $j_a\not\equiv j_b$ modulo 8. The remaining deconfined Ising pair splits into simpler Abelian components \begin{align}\Sigma^A_{j_a\pm1}\Sigma^B_{j_b\mp1}=S^+_{j_a\pm1,j_b\mp1}+S^-_{j_a\pm1,j_b\mp1},\end{align} where each $S^\pm$ carries the same spin as the original Ising pair but differs from the other by a unit fermion $S^\pm\times\Psi^{A/B}=S^\mp$. In general the two Abelian components are non-local with respect to each other. For instance, the \TR symmetric surface anyons $S^+$ and $S^-$ are mutually semionic when $j_a=j_b=0$. We choose to include $S^+$ in the condensate $b$ in \eqref{bosons} while confining $S^-$. Furthermore, the condensate identifies the deconfined anyons that are different up to bosons in $b$. \begin{align}\openone^A_{j_a}\openone^B_{j_b}&\equiv\Psi^A_{j_a}\Psi^B_{j_b}\equiv\Psi^A_{j_a+2}\openone^B_{j_b-2}\equiv\openone^A_{j_a+2}\Psi^B_{j_b-2}\nonumber\\&\equiv S^\pm_{j_a\pm1,j_b\mp1}\equiv\openone^A_{j_a+4}\openone^B_{j_b-4}\label{bulk}\end{align} for $j_a\equiv j_b$ mod 8 and $j_a,j_b$ both even. Equation~\eqref{bulk} are just parton combinations. For instance, $\psi^A=\Psi^A_4\openone^B_0\equiv\openone^A_4\Psi^B_4=\psi^B$ are now free to move inside both \FTI slabs after gluing. The \TO after the gluing is generated by the partons and dyons, which behave identically to those in $\mathcal{F}$ of \eqref{FTIFSFS}. This proves \eqref{gluing}. The anyon condensation gluing of the pair of \TPf states preserves symmetries for the same reason it does for the conventional \TI case~\cite{ChenFidkowskiVishwanath14,ChoTeoFradkin17}.

It is worth noting that a magnetic monopole can be mimicked by a magnetic flux tube / Dirac string (with flux quantum $hc/e$) that originates at the \TR symmetric surface interface and passes through one of the two \FTI slab, say the $A$ slab. In the prototype $a=-2$ and $g=-1$, the filling fraction $\nu_{\mathrm{Pf}^\ast}=1/2(2n+1)$ of the quasi-two-dimensional slab ensures, according to the Laughlin argument~\cite{Laughlin_IQHE}, that the monopole associates to the fractional charge $q=1/2(2n+1)$, which is carried by the confined \TPf surface anyons $\openone^A_2$ or $\Psi^A_2$. This surface condensation picture therefore provides a simple verification of the Witten effect~\cite{Witten79} for $\theta=\pi/(2n+1)$. 

Lastly, we noticed that in the band insulator case for $n=0$, $\mathcal{F}$ in \eqref{FTIFSFS} reduces to the Chern insulator or the lowest Landau level (\hypertarget{LLL}{LLL}), and $\mathrm{Pf}^\ast$ in \eqref{ZfTO} is simply the particle-hole (\hypertarget{PH}{PH}) symmetric Pfaffian state~\cite{Son15,BarkeshliMulliganFisher15,WangSenthil16}. The \PH symmetry is captured by the relative tensor product \eqref{gluing}, which can be formally rewritten into \begin{align}\mathrm{Pf}^\ast=\mathcal{F}\boxtimes\overline{\mathrm{Pf}^\ast}\label{gluing2}\end{align} by putting $\mathrm{Pf}^\ast$ on the other side of the equation. Here, the tensor product is relative with respect to some collection of condensed bosonic pairs, and $\overline{\mathrm{Pf}^\ast}$ is the \TR conjugate of $\mathrm{Pf}^\ast$. Equation~\eqref{gluing2} thus equates $\mathrm{Pf}^\ast$ with its \PH conjugate, which is obtained by subtracting itself from the \LLL. In the fractional case with $n>0$, \eqref{gluing2} suggests a generalized \PH symmetry for $\mathrm{Pf}^\ast$, whose \PH conjugate is the subtraction of itself from the \FQH state $\mathcal{F}$.

\section{\label{Conclusion}Conclusion}
To conclude, we studied gapped \FTI surface states with (i) \TR breaking order, (ii) charge $U(1)$ breaking order, as well as (iii) symmetry preserving \TPf topological order. We focused on \FTI that supported fractionally charged partons coupling with a discrete $\mathbb{Z}_{2n+1}$ gauge theory. We characterized the fractional interface channels sandwiched between different gapped surface domains by describing their charge and energy response, namely the differential electric and thermal conductance. The low-energy \CFT for these fractional interface channels corresponded to the \TO of quasi-$(2+1)$-D \FTI slabs with the corresponding gapped top and bottom surfaces. In particular, a \FTI slab with \TR conjugate ferromagnetic surfaces behaved like a fractional Chern insulator with \TO \eqref{FTIFSFS}, and in the particular case when $a=-1$ and $g=-2$, its charge sector was identical to that of the Laughlin $\nu=1/(2n+1)$ \FQH state. Combining the \TR symmetric \TPf surface with the \FTI bulk as well as the opposite \TR breaking surface, this \FTI slab exhibited a generalized Pfaffian \TO \eqref{ZfTO}. Furthermore, we demonstrated the gluing of a pair of parallel \TPf surfaces, which are supported by two \FTI on both sides. It was captured by an anyon condensation picture that killed the \TPf \TO and left behind deconfined partons and confined gauge and magnetic monopoles in the bulk. 

In Ref.~\cite{ChoTeoFradkin17} we also construct the \TPf state of the \FTI from the field theoretic duality approach.

\section*{Acknowledgments} 
J.C.Y.T. is supported by NSF Grant No.~DMR-1653535. G.Y.C. acknowledges the support from Korea Institute for Advanced Study (KIAS) grant funded by the Korea government (MSIP) and Grant No.~2016R1A5A1008184 under NRF of Korea.

\appendix

\section{Abelian Chern-Simons theory of dyons}
The fractional topological insulator slab with time-reversal conjugate surfaces has anyons which are dyons and partons. The neutral sector consists of only dyons. A dyon $\gamma$ is composed of $a$ number of $\mathbb{Z}_{2n+1}$ gauge charge on each surface associated with an unit gauge flux through the bulk. The dyons $\gamma^m$ where $m=0,1,\ldots,2n$, with $1=\gamma^0$ being the vacuum, form the anyon content of an Abelian topological state denoted as $\mathbb{Z}^{(2a)}_{2n+1}$. They have spins $h_{\gamma^m}=\frac{2am^2}{2n+1}$ modulo 1 and satisfy the $\mathbb{Z}_{2n+1}$ fusion rule $\gamma^m\times\gamma^{m'}=\gamma^{[m+m']}$, where $[m+m']$ is the remainder between 0 and $2n$ when dividing $m+m'$ by $2n+1$. For the case when $a=-1$, the Abelian topological theory becomes $\mathbb{Z}^{(-2)}_{2n+1}$, which is actually identical to $\mathbb{Z}^{(n)}_{2n+1}$. This is because the dyon ${\bf e}=\gamma^n$ has spin $\frac{-2n^2}{2n+1}\equiv\frac{n}{2n+1}$ modulo 1. The collection $\{{\bf e}^l:l=0,1,\ldots,2n\}$ is of 1-1 correspondence with $\{\gamma^m:m=0,1,\ldots,2n\}$. For instance $\gamma={\bf e}^{-2}={\bf e}^{2n-1}$. At the same time, $\mathbb{Z}^{(n)}_{2n+1}=\{{\bf e}^l:l=0,1,\ldots,2n\}$ is the anyon content of the Abelian Chern-Simons $SU(2n+1)_1$ theory with Lagrangian density $\mathcal{L}_{2+1}=\frac{1}{4\pi}\int_{2+1}K_{IJ}\alpha^I\wedge d\alpha^J$, where $\alpha^I$ for $I=1,\ldots,2n$ are $U(1)$ gauge fields, and \begin{align}K_{SU(2n+1)}=\left(\begin{array}{*{20}c}2&-1&&&&\\-1&2&-1&&&\\&-1&2&&&\\&&&\ddots&&\\&&&&2&-1\\&&&&-1&2\end{array}\right)\end{align} is the Cartan matrix of $SU(2n+1)$. 

\section{Anyon Condensation}

Here we will elaborate how to glue the two TR symmetric surfaces of a pair of $\mathrm{Pf}^\ast$ FTI slabs and condense surface bosonic anyon pairs on the two \TPf surfaces. As before we take the decoupled tensor product of the anyons in two $\mathrm{Pf}^\ast$ TO, denoted  $(\mathrm{Pf}^\ast)^A\otimes(\mathrm{Pf}^\ast)^B$ where $A,B$ refers to the two slabs. Then we choose a set of bosons that braid trivially around each other. 

First notice that dyon combinations $\gamma^z\equiv \tilde\openone^A_{0,z}\tilde\openone^B_{0,z}$ are not confined. A particle with charge ``$j$'' has gauge charge $n^2gj$, so our neutral pairs have gauge charge $n^2gj \times -n^2gj$. Thus the braiding phase with these dyons is $zn^2gj-zn^2gj = 0$.

Our parton should continuously move from slab $A$ to slab $B$, so we should condense $\Psi^A_{4}\Psi^B_{-4}$, the parton creation annihilation operator. Anything that braids with it is confined. We can derive braiding statistics with the ribbon formula, $\theta_{A,B}=h_{A \times B}-h_A-h_B$. The braiding phase from the anyon combination $\tilde{X}^A_{j_a,z_a}\tilde{X}^B_{j_b,z_b}$ around $\Psi^A_{4}\Psi^B_{-4}$ is is the same as $(\delta^A)^{z_a+n^3ugj_a}(\delta^B)^{z_b+n^3ugj_b}$ around $\Psi^A_{4}\Psi^B_{-4}$. The parton carries ``$g$'' gauge charge so this phase is  $g(z_a+n^3ugj_a-z_b-n^3ugj_b)$. This is zero if the dyon number $z+n^3ugj$ is equal on the $A$ and $B$ particle. This ensures gauge fluxes must continue through both $A$ and $B$ slabs, i.e., confines gauge magnetic monopoles. This means that we are left with combinations ${X}^A_{j_a}{X}^B_{j_b}\gamma^z$. It also identifies $\Psi^A_{4}\Psi^B_{-4}$ with the vacuum, which identifies 
\
 \begin{align*}
	\openone^A_{j_a}\openone^B_{j_b}\gamma^z&\equiv\Psi^A_{j_a+4}\Psi^B_{j_b-4}\gamma^z \equiv\openone^A_{j_a+8}\openone^B_{j_b-8}\gamma^z, \\
	\openone^A_{j_a}\Psi^B_{j_b}\gamma^z&\equiv\Psi^A_{j_a+4}\openone^B_{j_b-4}\gamma^z \equiv\openone^A_{j_a+8}\Psi^B_{j_b-8}\gamma^z, \\
	\Sigma^A_{j_a}\Sigma^B_{j_b}\gamma^z&\equiv\Sigma^A_{j_a+4}\Sigma^B_{j_b-4}\gamma^z, \\
	\openone^A_{j_a}\Sigma^B_{j_b}\gamma^z&\equiv\Psi^A_{j_a+4}\Sigma^B_{j_b-4}\gamma^z \equiv\openone^A_{j_a+8}\Sigma^B_{j_b-8}\gamma^z\\
&\equiv\Psi^A_{j_a+12}\Sigma^B_{j_b-12}\gamma^z.
\end{align*}

Next we choose the fermion pair $\Psi_0^A \times \Psi^B_0$. Notice $\Sigma$ braids with $\Psi$, so anything with just one $\Sigma$ is confined. This brings the identification to

\begin{align*}
	&\openone^A_{j_a}\openone^B_{j_b}\gamma^z  \equiv\openone^A_{j_a+4j}\openone^B_{j_b-4j}\gamma^z  \equiv\Psi^A_{j_a+4j}\Psi^B_{j_b-4j}\gamma^z, \\
	&\openone^A_{j_a}\Psi^B_{j_b}\gamma^z  \equiv\openone^A_{j_a+4j}\Psi^B_{j_b-4j}\gamma^z \equiv\Psi^A_{j_a+4j}\openone^B_{j_b-4j}\gamma^z, \\
	&\Sigma^A_{j_a}\Sigma^B_{j_b}\gamma^z\equiv\Sigma^A_{j_a+4j}\Sigma^B_{j_b-4j}\gamma^z. 
\end{align*}

Next we can condense $\Psi^A_{2}\openone^B_{-2}$, which when braided around $\openone^A_{j_a}\openone^B_{j_b}$ or $\Psi^A_{j_a}\openone^B_{j_b}$  gives $4(j_a-j_b)/16$ which is not confined if $j_a-j_b=0$ mod 4. For $\Sigma^A_{j_a}\Sigma^B_{j_b}$  gives $4(j_a-j_b)/16+1/2$ which is not confined if $j_a-j_b=2$ mod 4. The identification is now

\begin{align*}
	\openone^A_{j_a}\openone^B_{j_b}\gamma^z&\equiv\openone^A_{j_a+4j}\openone^B_{j_b-4j}\gamma^z\equiv\Psi^A_{j_a+4j}\Psi^B_{j_b-4j}\gamma^z\\ &\equiv\openone^A_{j_a+2}\Psi^B_{j_b-2}\gamma^z\equiv\openone^A_{j_a+2+4j}\Psi^B_{j_b-2-4j}\gamma^z 
	\\&\equiv\Psi^A_{j_a+2+4j}\openone^B_{j_b-2-4j}\gamma^z, \\
	\Sigma^A_{j_a}\Sigma^B_{j_b}\gamma^z&\equiv\Sigma^A_{j_a+2j}\Sigma^B_{j_b-2j}\gamma^z.
\end{align*}

Our $\Sigma \Sigma$ pairs now split into simpler Abelian components
\
 \begin{align}\Sigma^A_{j_a\pm1}\Sigma^B_{j_b\mp1}=S^+_{j_a\pm1,j_b\mp1}+S^-_{j_a\pm1,j_b\mp1},\end{align}
\
\noindent
where each $S^\pm$ carries the same spin as the original Ising pair but differs from each other by a unit fermion $S^\pm\times\Psi^{A/B}=S^\mp$. $S^+$ and $S^-$ normally have non-trivial mutual monodromy. We choose to condense the electrically neutral $S^+_{1,-1}$ and its multiples, while confining  $S^-_{1,-1}$. This means $\Sigma^A_{1}\Sigma^B_{-1}$ is condensed. The $\Sigma$ pair around $\openone^A_{j_a}\openone^B_{j_b}$ gives a phase of $2(j_a-j_b)/16$ which is zero if $j_a-j_b=0$ mod 8. The $\Sigma$ pair around $\openone^A_{j_a}\Psi^B_{j_b}$ gives a phase of $2(j_a-j_b)/16+1/2$ which is zero if $j_a-j_b=4$ mod 8. The $\Sigma$ pair around $\Sigma^A_{j_a}\Sigma^B_{j_b}$ gives a phase of $2(j_a-j_b)/16\pm1/4$ which is zero if $j_a-j_b=2$ or $6$ mod 8.

This then completes the full condensate, and we have the identification
\begin{align}
     \openone^A_{j_a}\openone^B_{j_b}\gamma^z&\equiv\Psi^A_{j_a,z}\Psi^B_{j_b,z}\gamma^z\equiv\Psi^A_{j_a+2}\openone^B_{j_b-2}\gamma^z\nonumber\\&\equiv\openone^A_{j_a+2}\Psi^B_{j_b-2,z}\gamma^z\equiv S^\pm_{j_a\pm1,j_b\mp1}\gamma^z\nonumber\\&\equiv\openone^A_{j_a+4}\openone^B_{j_b-4}\gamma^z\label{bulk2}
\end{align}
for $j_a\equiv j_b$ mod 8 and $j_a,j_b$ both even. This ends up being just the multiples of the parton $\openone^A_{0}\Psi^B_{4}$ together with the dyons $\gamma^z$. Together they generate the theory $\mathcal{F}$ of a FTI slab with two conjugate TR breaking surfaces.


%

\end{document}